\documentclass[useAMS,usenatbib,usegraphicx]{mn2e}

\usepackage[T1]{fontenc}
\usepackage{lmodern}
\newcommand{\pirm}{\mathrm{\pi}}

\renewcommand{\vec}[1]{\bmath{#1}}
\newcommand{\dfrac}[2]{\frac{#1}{#2}}
\newcommand{\transpose}[1]{{#1}^{\rmn{T}}}
\newcommand{\mat}[1]{\bmath{\mathsf{#1}}}
\newcommand{\matr}[2]{\mat{#1}^{\rmn{row}}_{#2}}
\newcommand{\matc}[2]{\mat{#1}^{\rmn{col}}_{#2}}
\newcommand{\matt}[1]{\transpose{\mat{#1}}}
\newcommand{\matrt}[2]{\mat{#1}^{\transpose{\rmn{row}}}_{#2}}

\newcommand{\mati}{\mat{I}}
\newcommand{\lsqreq}{\stackrel{\chi^2}{=}}
\newcommand{\iprod}{\cdot}
\newcommand{\oprod}{\otimes}
\newcommand{\eprod}{\circ}
\newcommand{\bigo}[1]{\mathcal{O}\left(#1\right)}
\newcommand{\norm}[1]{\left\|#1\right\|}

\newcommand{\abs}[1]{\left|#1\right|}
\newcommand{\diag}[1]{\rmn{diag}\left(#1\right)}
\newcommand{\var}[1]{\sigma_{#1}^2}
\newcommand{\covar}[2]{\sigma_{#1,#2}^2}
\newcommand{\pair}[2]{\left\langle#1,#2\right\rangle}
\newcommand{\func}[2]{\mathcal{#1}\left({#2}\right)}
\newcommand{\eqref}[1]{(\ref{#1})}

\newcommand{\ntxt}[1]{{\textit{N}_{\rmn{#1}}}}
\newcommand{\chisqtxt}[1]{{\chi_{\rmn{#1}}^2}}
\newcommand{\sigtxt}[1]{{\sigma_{\rmn{#1}}}}
\newcommand{\ltxt}[1]{{\lambda_{\rmn{#1}}}}
\newcommand{\nobs}{{\ntxt{obs}}}
\newcommand{\nvar}{{\ntxt{var}}}
\newcommand{\ncomp}{{\ntxt{comp}}}
\newcommand{\niter}{{\ntxt{iter}}}
\newcommand{\nrefine}{{\ntxt{refine}}}
\newcommand{\nbad}{{\ntxt{bad}}}
\newcommand{\sigin}{{\sigtxt{in}}}
\newcommand{\nset}{{\ntxt{set}}}
\newcommand{\chisqfit}{{\chisqtxt{fit}}}
\newcommand{\chisqtest}{{\chisqtxt{test}}}

\newcommand{\lrest}{\ltxt{rest}}

\newcommand\ion[3]{#1$\;${\small\rmfamily{#2}}$\lambda${#3}\relax}%

\begin{document}

\title[Weighted eigendecomposition approach to PCA]{Weighted principal component analysis:
a weighted covariance eigendecomposition approach}

\author[L. Delchambre]{L. Delchambre$^{1}$\thanks{E-mail:	ldelchambre@ulg.ac.be} \\
$^{1}$ Institut d'Astrophysique et de G\'eophysique, Universit\'e de Li\`ege, All\'ee du 6 Ao\^ut 17, B-4000 Sart Tilman (Li\`ege), Belgium}

\date{Accepted 2014 October 12. Received 2014 October 7; in original form 2014 July 4}

\volume{446(2)}
\pagerange{3545--3555}
\pubyear{2014}

\maketitle

\label{firstpage}

\begin{abstract}
We present a new straightforward principal component analysis (PCA) method based on the diagonalization of the weighted variance--covariance matrix through two spectral decomposition methods: power iteration and Rayleigh quotient iteration. This method allows one to retrieve a given number of orthogonal principal components amongst the most meaningful ones for the case of problems with weighted and/or missing data. Principal coefficients are then retrieved by fitting principal components to the data while providing the final decomposition. Tests performed on real and simulated cases show that our method is optimal in the identification of the most significant patterns within data sets. We illustrate the usefulness of this method by assessing its quality on the extrapolation of Sloan Digital Sky Survey quasar spectra from measured wavelengths to shorter and longer wavelengths. Our new algorithm also benefits from a fast and flexible implementation.
\end{abstract}

\begin{keywords}
  methods: data analysis -- quasars: general.
\end{keywords}

\section{Introduction}
\label{sec:intro}
	Principal component analysis (PCA) is a well-known technique initially designed to reduce the dimensionality of a typically huge data set while keeping most of its variance \citep{pearson1901, hotelling1933}. PCA is intimately related to the singular value decomposition (SVD) since the principal components of a data set, whose arithmetic mean is zero, will be equal to the eigenvectors of the covariance matrix sorted by their corresponding eigenvalue; or equivalently by the variance they account for. The principal coefficients are the linear coefficients allowing us to reconstruct the initial data set based on the principal components. Further details about PCA will be given in Section \ref{sec:pca} of this paper. Interested readers are also invited to read \cite{schlens2009} for an accessible tutorial on this technique or \cite{jolliffe2002} for a deeper analysis.

	PCA has many applications in a wide variety of astronomical domains from the classification of the Sloan Digital Sky Survey (SDSS) quasar spectra and their redshift determination \citep{yip2004,paris2014} to the study of the point spread function variation in lensing surveys \citep{jarvis2004}. 
The method described hereafter was originally developed in the framework of the \textit{Gaia} astrophysical parameters inference system \citep{cbj2013} where it is used to provide learning data sets of spectrophotometric data based on SDSS quasar catalog spectra \citep{paris2014}. The latter cover the observed wavelength range 4000--10 000\r{A} and are extrapolated by our algorithm to the wavelength range 3000--11 000\r{A} covered by \textit{Gaia}. Even if developed for an astronomical purpose, it can be used in any problems requiring PCA decomposition of weighted data. The case of missing data being simply the limiting case of weights equal to zero.
 
	Classical PCA is a mature tool whose performance in dimensionality reduction and pattern recognition has been assessed for a long time. Nevertheless, its main limitation comes from the fact that it is not adapted to the case of samples having weighted and/or missing data. The inherent consequence is that the classical PCA implementations made no difference between variance coming from a genuine underlying signal and variance coming from measurement noise.

	Most of the previous works cope with these limitations mainly by focusing on bypasses to the problem of noisy and/or missing data; or deal explicitly with particular cases. These encompass, for example, the interpolation of missing data \citep{beale1975} or cases where the weight matrix can be factorized into per-observation and per-variable weight matrices \citep{greenacre1984}. \cite{jolliffe2002} in sections 13.6 and 14.2 makes the point about these proposed solutions.
	
	At the present time, some methods are still able to deal with weight matrices having the same size as the corresponding data set \citep{gabriel1979,wentzell1997,tipping1999,srebro2003}. Nevertheless, none of these are able to provide the orthogonal principal components ordered by the data set variance it accounts for. Rather, they provide an unsorted set of not-necessary orthogonal vectors whose linear combination is optimized to describe the underlying variance but whose goal is not compatible with the explanation of the variance given a minimal number of vectors.

	Based on the idea of \cite{roweis1997}, \cite{bailey2012} and \cite{tsalmantza2012} have recently proposed methods based upon the expectation-maximization (EM) algorithm \citep{dempster1977} in order for the PCA to include weights associated with each variable within each observation. The objective of these methods is globally similar to the one of this paper. Differences mainly come from the fact that \cite{tsalmantza2012} aim at finding an orthogonal decomposition of the original matrix such that the whole data set variance is the best accounted for. Instead, our implementation focuses on finding the orthogonal vectors that are the best at individually describing the data set variance at the expense of a lower explained global variance. This trade-off comes from the fact that in a weighted case, the solution to the problem of finding the set of $N$ components explaining most of the variance of a data set is not guaranteed to contain the eigenvectors that are the best at individually describing this variance.  The implementation of \cite{bailey2012} takes benefits of the flexibility of the EM algorithm in order to interpolate between these two solutions.

	In Section \ref{sec:notation}, we explain the notation used in this paper. We summarize the properties of the classical PCA in Section \ref{sec:pca}. We see in details two current alternative implementations of weighted PCAs in Section \ref{sec:WEMPCA}. In Section \ref{sec:WPCA_new}, we describe our new algorithm while in Section \ref{sec:comparison}, we see its application on simulated data and real cases and compare it against other algorithms. Finally, some properties and extensions are discussed in Section \ref{sec:discussion} and we conclude in Section \ref{sec:conclusion}.

\section{Notation}
\label{sec:notation}
This paper uses the following notations: vectors are in bold italic, $\vec{x}$; $x_i$ being the element $i$ of the vector $\vec{x}$. Matrices are in uppercase boldface or are explicitly stated; i.e. $\mat{X}$ from which the $i$th row will be denoted $\matr{X}{i}$ and the $j$th column by $\matc{X}{j}$, element at row $i$, column $j$ will then be $\mat{X}_{ij}$. Amongst matrix operators, $\mat{a} \eprod \mat{b}$ denotes the element-wise product (Hadamard product) of $\mat{a}$ and $\mat{b}$ and $\norm{\mat{a}}$ denotes the Euclidian matrix norm of $\mat{a}$.

	Consider a problem where we have $\nobs$ observations each containing $\nvar$ variables, from which we want to retrieve $\ncomp$ principal components. For reference, here are the often used matrices along with their corresponding sizes: $\mat{X}$ the data set matrix ($\nvar \times \nobs$) from which we have subtracted the mean observation $\mat{\bar{y}}$ ($\matc{X}{i} = \matc{Y}{i} - \mat{\bar{y}}$); $\mat{W}$ the weight of each variable within each observation ($\nvar \times \nobs$); $\mat{P}$ the orthogonal matrix of principal components ($\nvar \times \ncomp$); $\matc{P}{i}$ being the $i$th principal component; $\mat{C}$ the principal coefficient matrix ($\ncomp \times \nobs$); $\mat{\sigma^2}$ the symmetric matrix of variance--covariance ($\nvar \times \nvar$) associated with $\mat{X}$.
	
	Finally, $\mat{A} \lsqreq \mat{B}$ means that $\mat{A}$ is the nearest matrix from $\mat{B}$ in a -- potentially weighted -- least-squares sense. Mathematically, this is equivalent to have $\mat{A}$ and $\mat{B}$ such that
\begin{equation}
\chi^2 = \sum_{ij} \mat{W}^2_{ij} \left(\mat{A}_{ij} - \mat{B}_{ij}\right)^2
\end{equation}
is minimized.

\section{Principal Component Analysis}
\label{sec:pca}
	Regarding classical PCA and assuming -- without any loss of generality -- that we would like to retrieve as many principal components as the number of variables (ie. $\ncomp = \nvar$),  then the goal of the PCA will be to find a decomposition 
\begin{equation}
\label{eq:pca_x_pc}
\mat{X} = \mat{P} \mat{C},
\end{equation}
such that 
\begin{equation}
\label{eq:pca_d_pvp}
\mat{D} = \matt{P} \mat{\sigma}^2 \mat{P} = \matt{P} \mat{X} \matt{X} \mat{P}
\end{equation}
is diagonal and for which
\begin{equation}
\label{eq:pca_d_ordered}
\mat{D}_{ii} \geq \mat{D}_{jj}; \; \forall i < j.
\end{equation}
	Note that based on equation \eqref{eq:pca_d_pvp} and according to the spectral theorem\footnote{Any real symmetric matrix is diagonalized by a matrix of its eigenvectors.}, $\mat{P}$ will be orthogonal.

\begin{figure}
\includegraphics{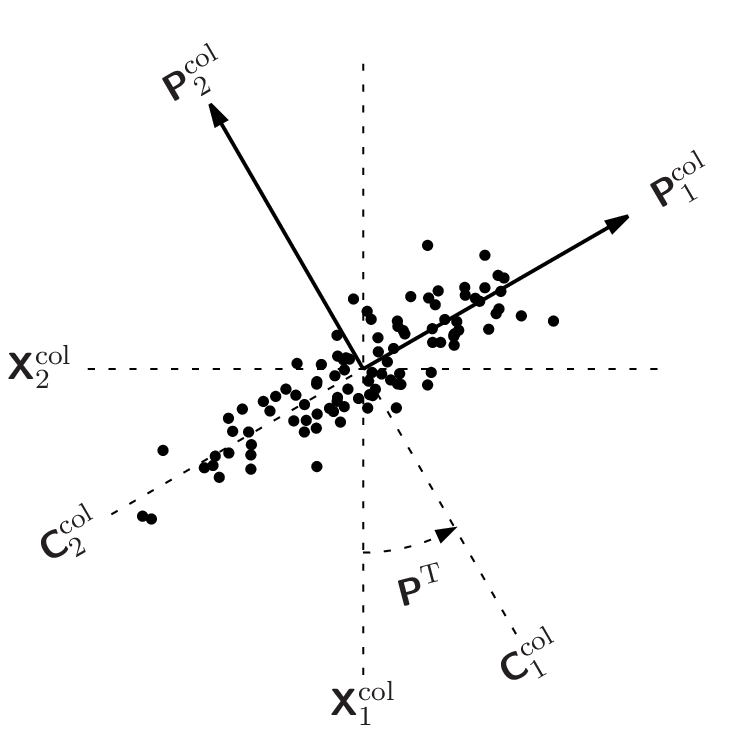}
\caption{A two-dimensional PCA example: $\matt{P}$ can be seen as an orthogonal coordinate transformation from $\left( \mat{X}^{\rmn{col}}_{1}; \mat{X}^{\rmn{col}}_{2} \right)$ to $\left( \mat{C}^{\rmn{col}}_{1}; \mat{C}^{\rmn{col}}_{2} \right)$ such that the data set variance is maximized along the vector $\matc{P}{1}$ and $\matc{P}{2}$. Note that, for a didactical purpose, we chose $\mat{P}$ to be a rotation matrix but practically it can be any orthogonal matrix.}
\label{fig:pca_example}
\end{figure}

	Intuitively, the matrix $\mat{P}$ can be seen as a change of basis allowing us to maximize the variance within $\mat{D}$ and thus minimizing the off-diagonal elements corresponding to the covariance. 	Differently stated, each $\matc{P}{i}$ defines a privileged direction along which the data set variance is the best explained. The fact that $\mat{D}$ is ordered implies that for $i < j$, the principal component $\matc{P}{i}$ accounts for more -- or equal -- variance than $\matc{P}{j}$. For the sake of clarity, a comprehensive PCA example is given in Fig. \ref{fig:pca_example}.
		
	A common solution to such a classical PCA is based on the SVD of $\mat{X}$:
\begin{equation}
\mat{X} = \mat{U} \mat{\Sigma} \matt{V},
\end{equation}
	where $\mat{U}$, $\mat{V}$ are orthogonals, $\mat{\Sigma}$ is diagonal and for which $\abs{\mat{\Sigma}_{ii}} \geq \abs{\mat{\Sigma}_{jj}}; \; \forall i < j$. By setting $\mat{P} = \mat{U}$ and $\mat{C} = \mat{\Sigma} \matt{V}$, we find that equation \eqref{eq:pca_d_pvp} becomes
\begin{equation}
\label{eq:pca_cc_d}
\matt{P} \mat{X} \matt{X} \mat{P} =  \mat{C} \matt{C} = \mat{\Sigma}^2,
\end{equation}
that fulfils the conditions of equations \eqref{eq:pca_d_pvp} and \eqref{eq:pca_d_ordered}. Note that in equation \eqref{eq:pca_d_pvp} the exact variance--covariance matrix should be normalized by $\nobs$ but since we are solely interested in the diagonalization of $\mat{\sigma^2}$, we drop it.

\section{Weighted Expected Maximization PCA}
\label{sec:WEMPCA}
	As already mentioned in Section \ref{sec:intro}, the current methods efficiently dealing with the problem of weighting PCA \citep{bailey2012,tsalmantza2012} aim at best explaining the whole data set variance according to a given number of principal components, which is equivalent to minimize
\begin{equation}
\label{eq:WPCA_x_pc}
\chi^2 = \sum_{ij} \mat{W}_{ij}^2 \left(\mat{X}_{ij} - \left[\mat{P} \mat{C} \right]_{ij} \right)^2,
\end{equation}
where the weighted mean observation we subtracted (cf. Section \ref{sec:notation}) is given by 
\begin{equation}
\mat{\bar{y}} = \sum_i \matc{W}{i} \matc{Y}{i} / \sum_i \matc{W}{i}.
\end{equation}
	We notice that equation \eqref{eq:WPCA_x_pc} has latent variables such that it has to rely on an iterative procedure to be solved.

	The EM algorithm is a statistical tool specifically designed to optimize a likelihood function for models having latent -- or unknown/hidden -- variables \citep{dempster1977}. This iterative procedure is composed of two steps.
\begin{enumerate}
\item E-step: find the expectation value of the latent variables given the parameters of the current model.
\item M-step: find the parameters of the model such that the likelihood function is optimized.
\end{enumerate}

	Based on the latter, \cite{roweis1997} has developed a fast and straightforward classical PCA algorithm for which the conditions in equations \eqref{eq:pca_x_pc}, \eqref{eq:pca_d_pvp} and \eqref{eq:pca_d_ordered} are all fulfilled. Regarding weighted PCA and more specifically the $\chi^2$ described by equation \eqref{eq:WPCA_x_pc}, we will have the following weighted expected maximization PCA (WEMPCA) algorithm:
\begin{description}
\item $\mat{P} \longleftarrow$ Random orthogonal matrix
\item While $\mat{P}$ and $\mat{C}$ have not converged
\item \quad\textit{(E-step)}~Find $\mat{C}$ that minimizes $\chi^2$ given $\mat{P}$.
\item \quad\textit{(M-step)}~Find $\mat{P}$ that minimizes $\chi^2$ given $\mat{C}$.
\end{description}
Note that the convergence criterion is still relative. It can be based on the $\chi^2$ -- or the change in the principal components $\Delta \mat{P}$ -- falling under a given threshold, the fact that the algorithm has reached a given number of iterations or whatever criterion we consider as relevant.

\subsection{Tsalmantza's implementation}
\label{sec:tsalmantza}
	\cite{tsalmantza2012} designed a general approach to the modelling and dimensionality reduction of the SDSS spectra called `Heteroskedastic Matrix Factorization'. More specifically, it attempts to minimize
\begin{equation}
\label{eq:tslamantza_chi2}
\chi_{\epsilon}^2 = \chi^2 + \epsilon \sum_{i > 1} \sum_j \left[ \mat{P}_{ij} - \mat{P}_{(i-1)j} \right]^2,
\end{equation}
subject to
\begin{equation}
\label{eq:tslamantza_constraints}
\left.
\begin{array}{l}
\mat{P}_{ij} \geq 0 \\
\mat{C}_{jk} \geq 0
\end{array} \right\rbrace \; \forall i, j, k.
\end{equation}
We recognize the first part of equation \eqref{eq:tslamantza_chi2} as being equation \eqref{eq:WPCA_x_pc} while the second part is a smoothing regularization term whose scalar $\epsilon$ defines the strength. Non-negativity constraints reflect a particular need to have a meaningful physical interpretation of the resulting spectra.

	Regarding the fact that we would like to model the widest variety of data sets, we will drop the non-negativity constraints that otherwise would have restricted our search space. Concerning the smoothing regularization factor, we have to note that it will be highly problem-dependent and that it can be tricky to optimize, this will result in a potential unfair comparison with other methods. We will then consider the case $\epsilon = 0$. Moreover, as we will see in Section \ref{sec:discussion_smoothing}, our method can deal with principal components smoothing as well, consequently ignoring it will not constitute a major drawback to our implementation.

	The resulting function to optimize will then be reduced to the sole equation \eqref{eq:WPCA_x_pc}. Nevertheless, and before going further, we have to note that minimizing equation \eqref{eq:WPCA_x_pc} will provide us a lower-rank  matrix approximation of  $\mat{X}$ but it is not a sufficient condition for the resulting matrices $\mat{P}$ and $\mat{C}$ to be considered as a PCA decomposition. According to \cite{tsalmantza2012}, the solution to this problem can be solved in two steps. 
	
	First consider a lower-rank matrix decomposition of $\mat{X}$, similar to the one produced by the solution of equation \eqref{eq:WPCA_x_pc},
\begin{equation}
\mat{X} \lsqreq \mat{A} \mat{B},
\end{equation}
where for clarity, the sizes of these matrices are $\mat{A}(\nvar \times \ncomp)$ and $\mat{B}(\ncomp \times \nobs)$. Now suppose an orthogonal basis $\mat{P_0}$ of $\mat{A}$; such an orthogonal basis always exists for full-rank matrices and can be retrieved through a straightforward Gram--Schmidt process for example. The associated coefficients matrix $\mat{C_0}$ is then directly retrieved by
\begin{equation}
\mat{C_0} = \matt{P_0} \mat{A} \mat{B}.
\end{equation}

	Secondly, in a way similar to equation \eqref{eq:pca_cc_d}, we will take the classical PCA decomposition of $\mat{C_0}$,
\begin{equation}
\mat{C_0} = \mat{P_c} \mat{C},
\end{equation}
such that $\mat{C} \matt{C}$ is diagonal. The resulting principal coefficient matrix will then be given by
\begin{equation}
\mat{P} = \mat{P_0} \mat{P_c},
\end{equation}
that will be orthogonal and that will provide us with the final decomposition
\begin{equation}
\label{eq:WEMPCA_x_pc}
\mat{X} \lsqreq \mat{P} \mat{C}.
\end{equation}

	The above-mentioned steps have thus to be performed after the EM algorithm minimizing equation \eqref{eq:WPCA_x_pc} in order for $\mat{P}$ to be orthogonal and for the covariance matrix $\mat{C}\matt{C}$ to be diagonal. Sections \ref{sec:wempca_tsalmantza_estep} and \ref{sec:wempca_tsalmantza_mstep} will now focus on details of the EM algorithm.

\subsubsection{E-step}
\label{sec:wempca_tsalmantza_estep}
	
	As stated at the beginning of this section, the expectation step regarding WEMPCA will be given by the retrieval of the coefficient matrix $\mat{C}$ that minimizes equation \eqref{eq:WPCA_x_pc} -- or equivalently that optimizes equation \eqref{eq:WEMPCA_x_pc} -- while considering the principal component matrix $\mat{P}$ being held fixed. Since each observation -- column of $\mat{X}$ -- is a linear combination of the principal components, finding the solution of equation \eqref{eq:WEMPCA_x_pc} is equivalent to solving
\begin{equation}
\label{eq:wempca_xcol_pccol}
\matc{X}{i} \lsqreq \mat{P} \matc{C}{i}; \; \forall i	,
\end{equation}
whose solutions are given by the `Normal Equations':\footnote{This method is known to suffer from numerical instabilities \citep{press2002} and is provided for a didactical purpose only. Methods such as SVD for linear least squares must be preferred in order to solve equation \eqref{eq:wempca_xcol_pccol}.}
\begin{equation}
\label{eq:wempca_e_step}
\matc{C}{i} = \left(\matt{P} \mat{w}^2 \mat{P} \right)^{-1} \matt{P} \mat{w}^2 \matc{X}{i},
\end{equation}
with $\mat{w} = \diag{\matc{W}{i}}$.

\subsubsection{M-step}
\label{sec:wempca_tsalmantza_mstep}
	Similarly to the E-step, solution to the M-step -- that is the retrieval of $\mat{P}$ that optimizes equation \eqref{eq:WEMPCA_x_pc} given $\mat{C}$ -- can be decomposed by noting that within each observation, a given variable is the linear combination of the corresponding principal components variables. That is,
\begin{equation}
\label{eq:wempca_tsalmantza_mstep_objective}
\matr{X}{i} \lsqreq \matr{P}{i} \mat{C}; \; \forall i,
\end{equation}
whose solutions are
\begin{equation}
\label{eq:wempca_tsalmantza_mstep_solution}
\matr{P}{i} = \matr{X}{i} \mat{w}^2 \matt{C} \left(\mat{C} \mat{w}^2 \matt{C}\right)^{-1},
\end{equation}
with $\mat{w} = \diag{\matr{W}{i}}$.

\subsection{Bailey's implementation}
	As we have seen in Section \ref{sec:tsalmantza}, the implementation of \cite{tsalmantza2012} focuses on the solution of equation \eqref{eq:WPCA_x_pc} while the PCA decomposition can be seen as a supplemental step that is external to the EM algorithm. Moreover, a single iteration of the algorithm requires the solution of $\nobs + \nvar$ systems of linear equations, each of size ($\ncomp \times \ncomp$), at each iteration of the EM algorithm. This becomes quickly unmanageable regarding huge data sets.
	
	\cite{bailey2012} takes the pragmatic approach that the sole solution of equation \eqref{eq:WPCA_x_pc} can lead to good insights about the principal components if the latter were fitted individually. This hypothesis is reasonable since this will allow each individual principal component to maximize the variance it accounts for.
	
	The resulting implementation will be similar to the one of \cite{tsalmantza2012} apart from the optimization function of the M-step. Indeed, in order for the principal components to be fitted separately, we have to consider the cross-product decomposition of equation \eqref{eq:WEMPCA_x_pc}, that is
\begin{equation}
\label{eq:wempca_bailey_mstep}
\mat{X} \lsqreq \sum_j \matc{P}{j} \matr{C}{j},
\end{equation}
from which each $\matc{P}{j}$ has to be individually fitted.

	Suppose that we already retrieved the $(j-1)$ first principal components. Let us also assume that the data projection along these $(j-1)$ principal components was already subtracted from the data set, that is
\begin{equation}
\label{eq:wempca_bailey_proj_removal}
\mat{X}^\prime = \mat{X} - \sum_i^{j-1} \matc{P}{i} \matr{C}{i}.
\end{equation}	

	Then the retrieval of $\matc{P}{j}$ based on equation \eqref{eq:wempca_bailey_mstep} can be decomposed in a way similar to equation \eqref{eq:wempca_tsalmantza_mstep_objective} as
\begin{equation}
\mat{X}^{\prime\rmn{row}}_{i} \lsqreq \mat{P}_{ij} \matr{C}{j},
\end{equation}
whose solution is straightly given by
\begin{equation}
\mat{P}_{ij} = \dfrac{\mat{X}^{\prime\rmn{row}}_{i} \mat{w}^2 \matrt{C}{j}}{\matr{C}{j} \mat{w}^2 \matrt{C}{j}},
\end{equation}
with $\mat{w} = \diag{\matr{W}{i}}$.

	Equation \eqref{eq:wempca_bailey_proj_removal} theoretically ensures that the last-retrieved component, $\matc{P}{j}$ will be orthogonal to any previous one. Nevertheless, due to machine round-off errors, this has to be manually checked.
	
	Finally, we have to note that solving equation \eqref{eq:wempca_bailey_mstep} will not minimize the global $\chi^2$ -- as defined by equation \eqref{eq:WPCA_x_pc} -- such that the algorithm has to rely on a last E-step at the end of the main EM algorithm.

\section{New implementation}
\label{sec:WPCA_new}
	Though both mentioned algorithms \citep{bailey2012, tsalmantza2012} correctly find lower-rank orthogonal decompositions that are suitable to explain the whole data set variance at best, none of them assures us that the retrieved principal components will be those that maximize the individual variance described by each of them. 
These principal components are then efficient at reconstructing the initial data set but are not the best at individually describing the underlying data structure.
	
	The basic idea of this new algorithm is to focus on the maximization of the weighted variance explained by each principal component through the diagonalization of the associated weighted covariance matrix. The resulting principal components will then be those that are the most significant -- under the assumption that the definition of the used weighted variance is relevant -- in identifying pattern within the data set even if their linear combination is not necessarily the best at explaining the total data set variance as described by equation \eqref{eq:WPCA_x_pc}.
	
	In the following, we will consider that the weighted variance of a given discrete variable $\vec{x}$ having weights $\vec{w}$ is given by
\begin{equation}
\label{eq:var_new}
\var{\vec{x}} = \dfrac{\sum_i w_i^2 \left(x_i - \bar{x} \right)^2}{\sum_i w_i^2},
\end{equation}
	where $\bar{x} = \sum_i w_i x_i / \sum_i w_i$ and with the convention that $0/0=0$. The latter can be straightforwardly extended to the definition of the weighted covariance between two discrete variables, $\vec{x}$ and $\vec{y}$, for which the weights are given by $\vec{w^x}$ and $\vec{w^y}$ respectively, that is
\begin{equation}
\label{eq:covar_new}
\covar{\vec{x}}{\vec{y}} = \dfrac{\sum_i \left(x_i - \bar{x} \right) w_i^x w_i^y \left( y_i - \bar{y} \right)}{\sum_i w_i^x w_i^y}.
\end{equation}

	Based on these definitions, we can write the weighted covariance matrix of a data set $\mat{X}$ with associated weights $\mat{W}$ as
\begin{equation}
\label{eq:covar_mat_new}
\mat{\sigma^2} = \dfrac{\left(\mat{X} \eprod \mat{W}\right)\transpose{\left(\mat{X} \eprod \mat{W}\right)}}{\mat{W}\matt{W}}.
\end{equation}

	We know, from the spectral theorem, that there exists an orthogonal matrix $\mat{P}$ such that $\mat{\sigma^2}$ is diagonalized, and consequently that equation \eqref{eq:pca_d_pvp} is  fulfilled (as well as equation \ref{eq:pca_d_ordered} if $\mat{P}$ is ordered accordingly). This matrix $\mat{P}$ will then constitute the principal components of our implementation.

\begin{figure}
\includegraphics{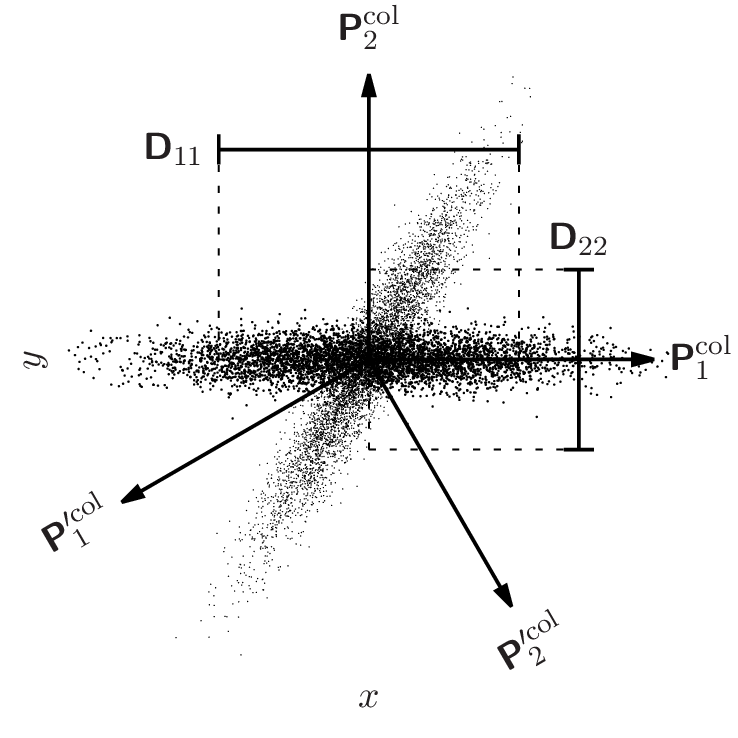}
\caption{Comparison between weighted principal components $\mat{P}$ and classical principal components $\mat{P^\prime}$ in presence of underweighted observations corresponding to the brighter points. $\abs{\mat{D}_{11}}$ and $\abs{\mat{D}_{22}}$ correspond to the variances respectively explained by $\matc{P}{1}$ and $\matc{P}{2}$.}
\label{fig:wpca_diff}
\end{figure}

	Fig. \ref{fig:wpca_diff} shows a two-dimensional example of classical principal components $\mat{P^\prime}$ that are unable to individually describe the underlying data variance. In this example, we have supposed -- for didactical purpose -- that the observations corresponding to the bright points have far lower weights associated with the $x$ variable. Weighted principal components, $\mat{P}$, diagonalizing $\mat{\sigma^2}$ as described by equation \eqref{eq:covar_mat_new} are given along with the variance it explains (that is the diagonal elements of $\mat{D} = \matt{P}\mat{\sigma}\mat{P}$). Note that such principal components maximize the variance explained by each variable as described by equation \eqref{eq:var_new} and consequently set to zero the associated covariance as described by equation \eqref{eq:covar_new}.
	
	The goal of the algorithm is then to retrieve the dominant eigenvector $\vec{p}$ from the covariance matrix $\mat{\sigma^2}$ along with its associated eigenvalue $\lambda$, that is the dominant eigenpair $\pair{\vec{p}}{\lambda}$. $\vec{p}$ will then be the principal component explaining most of the data set variance, $\lambda$. That is equivalent to find $\vec{p}$ in
\begin{equation}
\label{eq:eigenvector}
\mat{\sigma^2} \vec{p} = \lambda \vec{p},
\end{equation}
such that $\lambda$ is maximized\footnote{In fact, $\pair{-\vec{p}}{-\lambda}$ is also solution of equation \eqref{eq:eigenvector} but for the sake of clarity, we will only consider the case of positive eigenvalues}.

	Equation \eqref{eq:eigenvector}, corresponding to the eigenvector definition, is a well-studied problem for which many methods already exist. The reference in the domain is \cite{golub1996} where the interested reader may find a rather exhaustive list of such methods as well as proofs of the algorithms described hereafter. 
	
	Unsurprisingly, in the context or our implementation, we choose the fastest and simplest algorithm called the power iteration method. The idea behind this algorithm is to recognize that given a diagonalizable square matrix $\mat{A}$ and a vector $\vec{u}^{(0)}$ having a nonzero component in the direction of $\vec{p}$, the iterative relation
\begin{equation}
\label{eq:power_iteration}
\vec{u}^{(k)} = \mat{A} \vec{u}^{(k-1)} = \mat{A}^k \vec{u}^{(0)}
\end{equation}
will converge to a vector that is proportional to the dominant eigenvector $\vec{p}$ as $k \rightarrow \infty$. Note that in practice, each vector $\vec{u}^{(k)}$ is normalized to unity in order to avoid numerical round-off errors inherent to the computer representation of large numbers. The final eigenvector will then be given by $\vec{p} = \vec{u}^{(k)} / \norm{\vec{u}^{(k)}}$ and the associated eigenvalue by the Rayleigh quotient:
\begin{equation}
	\label{eq:rayleigh_quotient}
	\func{R}{\mat{A}, \vec{p}} = \vec{p} \iprod \mat{A}\vec{p} = \vec{p} \iprod \lambda\vec{p} = \lambda.
\end{equation}
 Convergence and assumptions made about this algorithm will be discussed in Section \ref{sec:discussion_convergence}. Further principal components can be retrieved by considering application of the above-mentioned algorithm to
\begin{equation}
\label{eq:rayleigh_subtract}
\mat{A^\prime} = \mat{A} - \lambda \vec{p} \oprod \vec{p},
\end{equation}
	that is the matrix obtained by subtracting the data variance along the found principal components.

\subsection{Refinement}
\label{sec:WPCA_new_refinement}
	As we will see in Section \ref{sec:discussion_convergence}, the power iteration method may have a slow convergence rate under some particular conditions. Consequently it may be that some vectors did not effectively converge to an eigenvector that would have diagonalized the covariance matrix. Nevertheless, \cite{parlett1974} proposed an algorithm -- called Rayleigh quotient iteration -- designed to tackle this kind of problem.
	
	Even if the proof of this algorithm is beyond the scope of this paper, we still mention two basic facts to enable the reader to have a minimal understanding of how it works. First, as we have seen in equation \eqref{eq:rayleigh_quotient}, the Rayleigh quotient of a matrix with one of its eigenvector is equal to its associated eigenvalue. Secondly, given a matrix $\mat{A}$ with eigenvalues $\lambda_1, ..., \lambda_n$, we find that the eigenvalues of the matrix $\left(\mat{A} - d \mati \right)^ {-1}$ will be $\left(\lambda_1 - d\right)^{-1}, ..., \left(\lambda_n - d\right)^{-1}$. Based upon these facts, we will have that the sequence
\begin{eqnarray}
\label{eq:rayleigh_iter_power}
\vec{u}^{(k)} & = & \left(\mat{A} - d^{(k-1)} \mati \right)^{-1} \vec{u}^{(k-1)} \\
d^{(k)} & = & \func{R}{\mat{A},\vec{u}^{(k)}},
\end{eqnarray}
where each $\vec{u}^{(k)}$ is normalized to the unit length, will converge cubically to the eigenpair $\pair{\vec{p}}{\lambda}$ that is the nearest -- regarding the absolute value of their Rayleigh quotient -- from a starting point $\pair{\vec{u}^{(0)}}{d^{(0)}}$ as $k \rightarrow \infty$. Note that equation \eqref{eq:rayleigh_iter_power} is a power iteration towards the eigenvector for which $\abs{\lambda - d^{(k-1)}}$ is minimized.

	Finally, the principal coefficients are retrieved by solving $\chi^2 \lsqreq \mat{P} \mat{C}$ whose `Normal Equation' solution is given by equation \eqref{eq:wempca_e_step}.
	
\subsection{Variance regularization}
	Real-world data often have sparse and unevenly distributed weights such that it may happen for some variables to have their corresponding variances to be based only on a small number of observations. Such situations may become problematic since these few variables will have a strong impact on the resulting first principal components in a way that is irrespective to their total weight.
	
	Such `overweighted' variables can be damped by using a regularization factor within the expression of the weighted covariance, $\covar{\vec{x}}{\vec{y}}$, as defined by equation \eqref{eq:covar_new}. The resulting regularized weighted covariance will take the form
\begin{equation}
\label{eq:covar_regularized}
\covar{\vec{x}}{\vec{y}}\left(\xi\right) = \left[ \sum_i w_i^x \sum_i w_i^y \right]^{\xi} \covar{\vec{x}}{\vec{y}},
\end{equation}
 where the regularization parameter $\xi$ allows us to control the damping strength.

	The typical value of the regularization parameter, $\xi$, goes from zero, where we get back to the classical behaviour of the algorithm, to two for a strong damping of these rare variables. Conversely one might want to highlight such underrepresented variables by setting the regularization parameter to a negative value.

\section{Comparison}
\label{sec:comparison}

	The performance of the method described in this paper was assessed regarding the two previously described algorithms, namely the one of \cite{bailey2012} and the one of \cite{tsalmantza2012}. The choice of these algorithms comes from the fact that they are fairly competitive and have goals that are comparable to those of the new algorithm. All methods were tested on both simulated data as well as on real observational ones.

\subsection{Simulated data}
\label{sec:comparison_simulated}
	Simulated data consist in random linear combinations of 10 orthogonal basis functions. These bases are produced by taking 10 shifted sine functions having periods between $0.2\pirm$ and $2\pirm$ and by applying a Gram--Schmidt orthogonalization process to the latter. Resulting observations are then sampled over 100 evenly spaced points in the interval $\left[0, 2\pirm\right]$. To each variable, $x$, within each observation, $\vec{x}$, we also add a Gaussian noise having a standard deviation given by
\begin{equation}
\sigma_x = \sigin \left( 1 + \sigtxt{obs} \right) \left(1 + \mathcal{U}_\sigma\right) \max \abs{\vec{x}},
\label{eq:simulation_sigma_x}
\end{equation}
where $\sigin$ is a user-provided parameter corresponding to the desired noise amplitude, $\sigtxt{obs}$ is an observation-specific noise ponderation uniformly drawn from $\left[-0.1, 0.1\right]$ and $\mathcal{U}_\sigma$ is a uniform random variable corresponding to the noise dispersion within observations and taking values in the range $\left[-0.1, 0.1\right]$. The weight associated with the variable $x$ will then be set to $1/\sigma_x$. 
	
\begin{figure}
\includegraphics{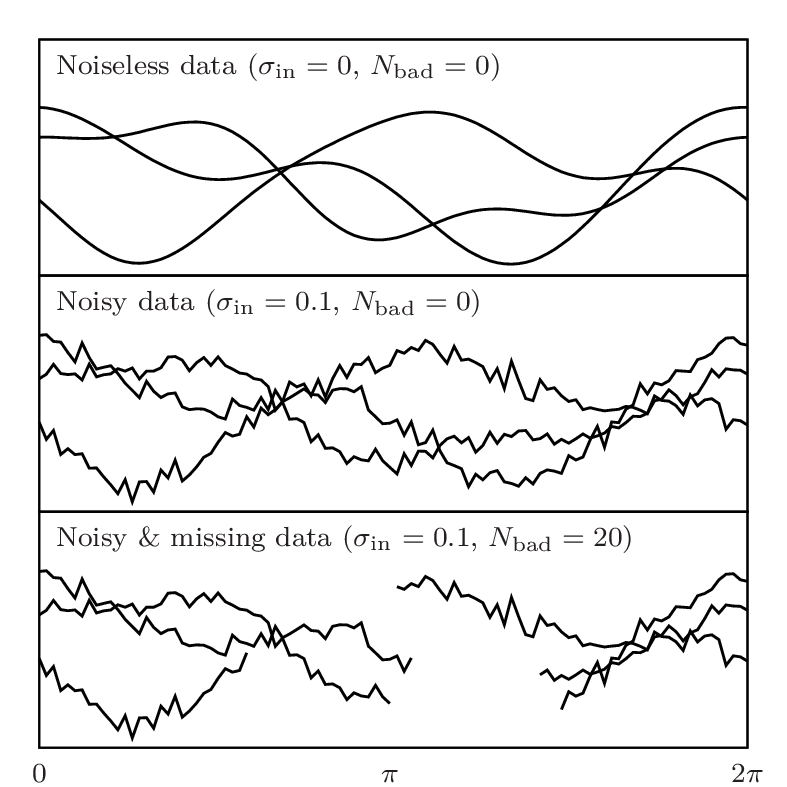}
\caption{Examples of simulated data illustrating the effect of the simulation parameters $\sigin$ and $\nbad$.}
\label{fig:comparison_example}
\end{figure}

	Finally, we discard $\nbad$ contiguous and randomly positioned variables from each observation. The latter will be used to assess performances of the various algorithms on data extrapolation while having their weights equal to zero during the PCA retrieval phase. Examples of such simulated data are illustrated in Fig. \ref{fig:comparison_example}.
	
	In order to perform the comparison with other algorithms, we built a given number of data sets, $\nset$, each containing 1000 observations. Each of these sets was duplicated and altered through the realization of various values of the simulation parameters $\sigin$ and $\nbad$. We then retrieved, for each of the mentioned algorithms, the five first principal components out of the resulting altered data sets and computed estimators based on the following $\chi^2$ definition:
\begin{equation}
\label{eq:data set_chisq}
\chi^2 = \dfrac{\sum_{ij} \left[ \mat{W} \eprod \left(\mat{X} - \mat{P} \mat{C} \right) \right]_{ij}^2}{\sum_{ij} \mat{W}_{ij}^ 2}.
\end{equation}	
	The following estimators were computed: $\chisqfit$, the chi-square of the data set for which weights associated with the discarded variables are set to zero and $\chisqtest$ where only rejected variables are considered and for which weights associated with the unrejected variables are set to zero. Let us note that $\chisqfit$ will typically account for the quality of the fit while $\chisqtest$ will account for the quality of the extrapolation. If these estimators are to be computed based on the number of data set, $\nset$, we use a $3\sigma$-clipped mean over all the $\chisqfit$ and $\chisqtest$ associated with each data set, namely $\langle\chisqfit\rangle$ and $\langle\chisqtest\rangle$.
	
	For completeness, only data sets having $\nbad \leq 50$ will be discussed here. This decision comes from the fact that efficiently estimating the principal components of such sparse data sets while having `only' 1000 observations is a really tricky task strongly depending on the design of these data sets. Furthermore, as we will see in Section \ref{sec:scaling_performance}, the Bailey and Tsalmantza algorithms are big time consumers such that dealing with bigger data sets quickly becomes unmanageable.

	Since all the studied algorithms are based on iterative procedures, we have to take into account the convergence criterion for each of them. To this aim, we performed a preliminary study whose goal is to determine the minimal number of iterations needed by each algorithm in order to reach convergence. This was assessed by running 100 times each algorithm on data sets similar to those previously described and by setting the initial eigenvectors estimates to random values within each run. In order to make sure we can model unseen and potentially more complex data sets, the number of iterations was set to twice the found number of iterations needed to converge, giving respectively $500$, $500$ and $10^4$ iterations for Bailey, Tsalmantza and our algorithm (without refinement).

\begin{figure}
\includegraphics{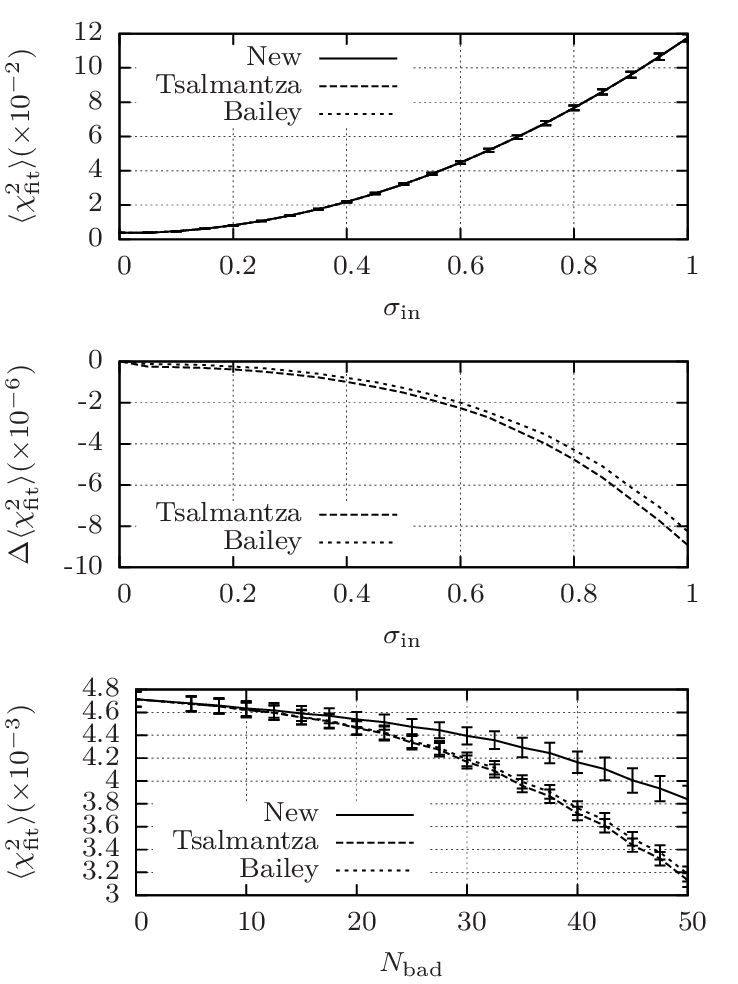}
\caption{Averaged $\chisqfit$ for increasing noise without missing data (top). Subtraction of Bailey and Tsalmantza $\chisqfit$ from our algorithm $\chisqfit$ (middle). Averaged $\chisqfit$ with moderate noise and an increasing number of missing data (bottom).}
\label{fig:comparison_chi2_fit}
\end{figure}
	
	Regarding the quality of the fit, $\chisqfit$, differences between the various algorithms are fairly low even if -- as expected -- our algorithm is proven to have somewhat larger $\chisqfit$ with a higher dispersion over all the data sets. Fig. \ref{fig:comparison_chi2_fit} shows the behaviour of the mean $\chisqfit$ regarding two common cases, namely the case of increasing noise and no missing data and the case of moderate noise ($\sigin = 0.1$) with increasing number of missing data. Each point on these graphs is averaged over $\nset = 1000$ data sets. Practically, one cannot distinguish the various algorithms if all data are present since differences are in $\bigo{10^{-6}}$. In presence of missing data, differences start to be noticeable but still reasonable with differences in $\bigo{10^{-3}}$. For a didactical purpose, let us note that if only the first component -- out of the five retrieved -- was considered then our algorithm would have had a better or equal $\chisqfit$ than Bailey and Tsalmantza implementations over all the values of the simulation parameters $\sigin$ and $\nbad$.

\begin{figure}
\includegraphics{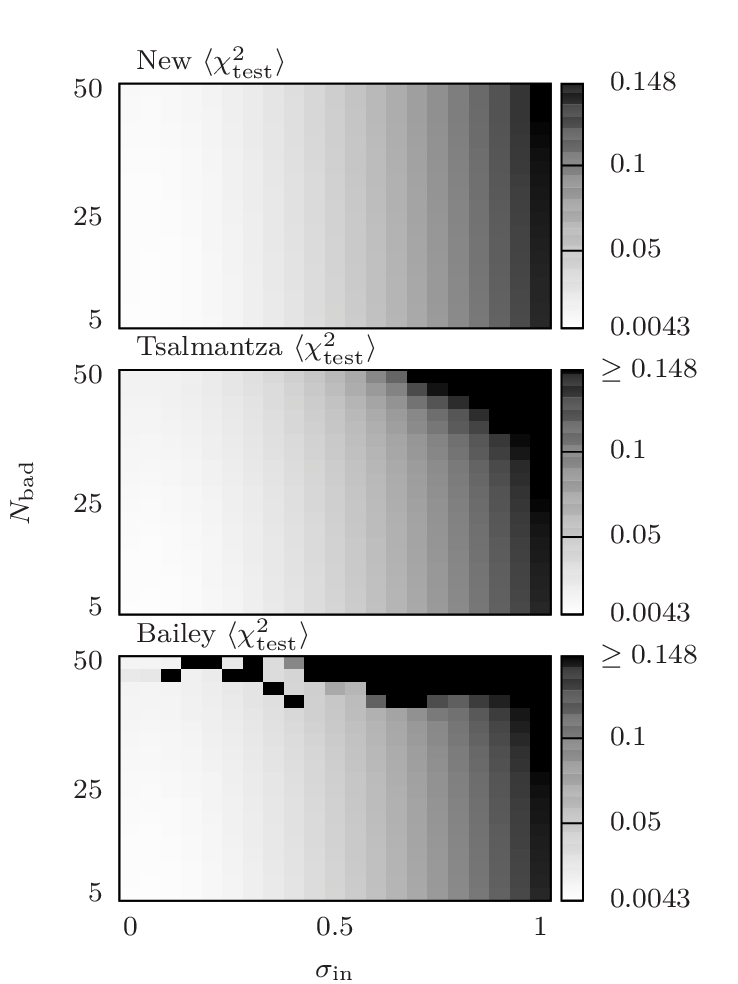}
\caption{$\chi^2$ maps of averaged $\chisqtest$ regarding $\sigin$ and $\nbad$ simulation parameters.}
\label{fig:comparison_chi2_test_map}
\end{figure}

\begin{figure}
\includegraphics{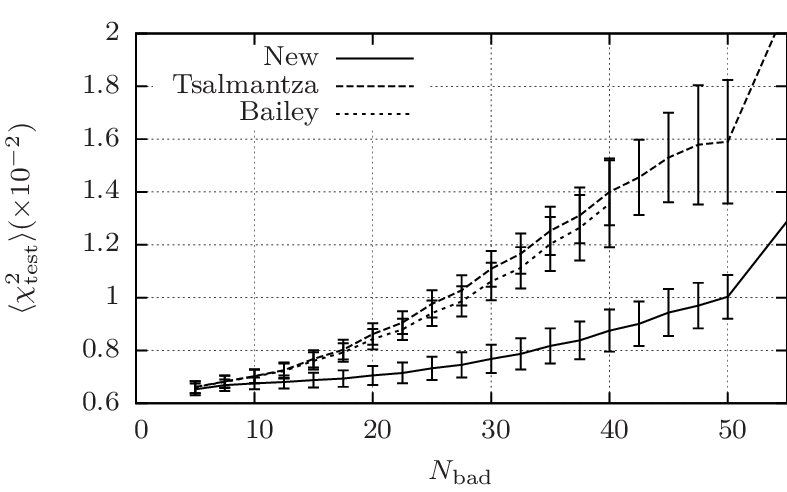}
\caption{Averaged $\chisqtest$ for moderate noise with an increasing number of missing data.}
\label{fig:comparison_chi2_test}
\end{figure}

	Regarding now the quality of the extrapolation, differences are more noticeable. Fig. \ref{fig:comparison_chi2_test_map} illustrates these discrepancies according to the averaged $\chisqtest$ over $\nset = 100$ data sets having reasonable $\chisqfit$ in each of the  $\sigin$ and $\nbad$ simulation parameters. At first glance, our new algorithm shows a globally better $\langle\chisqtest\rangle$ while suffering less from missing data: the latter being dominated by noise within the data. The Tsalmantza's algorithm shows good performances but has a stronger dependence on missing data, reaching a maximum $\langle\chisqtest\rangle = 0.372$ while Bailey's algorithm shows strong numerical instabilities for $\nbad > 40$ that makes it unable to converge and makes it reach a maximum $\langle\chisqtest\rangle \approx 3 \times 10^9$  for $\sigin = 0.9$ and $\nbad = 50$. Fig. \ref{fig:comparison_chi2_test} illustrates in more detail the behaviour of $\langle\chisqtest\rangle$ averaged over $\nset = 1000$ data sets in the common case of moderate noise ($\sigin = 0.1$) with an increasing number of missing data. For clarity, the plot of Bailey having $\nbad > 40$ has been removed from the graph.
	
	Finally, let us note that data sets having $\sigin = 0$ and $\nbad = 0$ can be solved using a classical PCA algorithm. Consequently, both the variance explained by each individual component and the total variance can be simultaneously optimized. We will thus find -- in this particular case -- that all algorithms will provide us with identical results as it was already suggested in Fig. \ref{fig:comparison_chi2_fit}. Similarly, if we choose to retrieve a single component, even with noisy and/or missing data, the algorithms of Bailey and Tsalmantza will maximize the variance explained by this component, the latter will thus match the first component that would have been retrieved by our algorithm.

\subsection{Observational data}
\label{sec:comparison_observation}

	Comparisons against a concrete case were performed using the SDSS DR10Q quasar catalog from \cite{paris2014}. Out of the 166 583 QSO spectra present in the initial data release, 18 533 were rejected either due to spectra bad quality, strong uncertainties in redshift determination, presence of BAL or insufficient number of high-S/N points. The remaining spectra were set to the rest frame; the continuum was then subtracted and the spectra were normalized such as to have a zero mean and a variance of 1. Finally, visual inspection showed that in some cases the continuum was badly fitted such that the variance within these spectra can mainly be attributed to this error; these regions were removed using a \textit{k}-sigma clipping algorithm for each variable among all observations. There remain 148 050 spectra having observed wavelengths between 4000 and 10 000\r{A} and for which the variance within each spectrum is thought to be mainly caused by genuine signals. In the following tests, the 10 first components will be retrieved from each algorithm. We also suppose, as previously, that $\xi = 0$.
	
	 The number of iterations associated with each algorithm was assessed by using a subset of the above-described data set (1000 $\leq \lrest \leq$ 2000\r{A}, $3 \leq z \leq 4$) and by running 10 times the various algorithms on it with random initial principal components. Convergence was assessed by minimizing the variance amongst the final principal components within these 10 runs. The fact that only a subset of the above-described data set was used can be explained by the large amount of time needed by each algorithm to run as well as by the fact that -- as we will see -- Tsalmantza and Bailey algorithms often fail to converge in presence of a large amount of missing data. The results show that the number of iterations chosen in Section \ref{sec:comparison_simulated} also match the complexity of this problem.

\begin{figure*}
\includegraphics{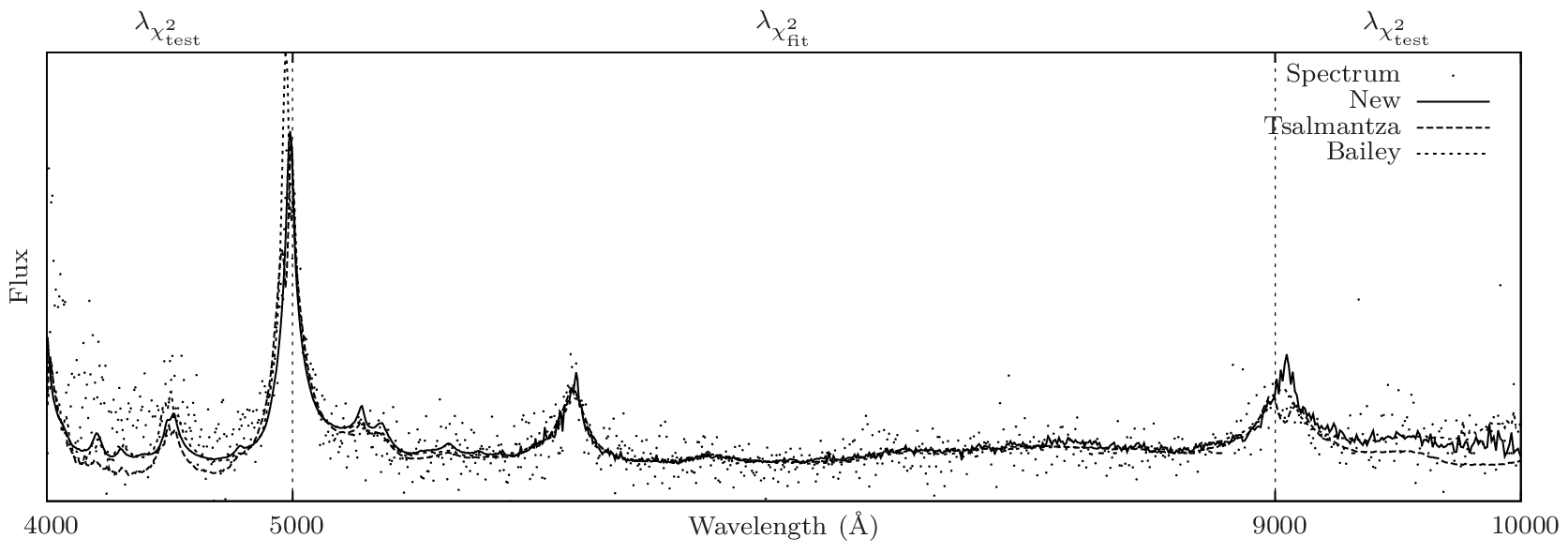}
\caption{Example of SDSS DR10Q spectrum fits. $\lambda_{\chisqtest}$ and $\lambda_{\chisqfit}$ denote regions used to compute, respectively, $\chisqtest$ and $\chisqfit$ for this observation.}
\label{fig:comparison_sdss_spec}
\end{figure*}
	
	The initial test to be performed is similar to tests performed in Section \ref{sec:comparison_simulated} in the sense that PCA were retrieved for all algorithms using the region 5000--9000\r{A} in observed wavelength while regions 4000--5000 and 9000--10 000\r{A} are rejected and kept to assess the quality of the extrapolation. Fig. \ref{fig:comparison_sdss_spec} shows an example of such a spectrum along with the fits by the various algorithms. The resulting SDSS DR10Q data set $\chisqfit$ and $\chisqtest$, as defined by equation \eqref{eq:data set_chisq}, are given in Table \ref{tbl:sdss_data set_chi2}.	

\begin{table}
\begin{tabular}{l|ccc}
\hline
 & New & Tsalmantza & Bailey \\
\hline
Data set $\chisqfit$ & 0.107 & 0.088 & 0.094 \\
Data set $\chisqtest$ & 1.064 & $2 \times 10^{5}$ & $8 \times 10^{12}$  \\
Median $\chisqtest$  & 1.021 & 1.789 & $8 \times 10^4$\\
Ratio of observations \\ having $\chisqtest \geq 5$ & 0.014 & 0.33 & 0.81 \\
\hline
\end{tabular}
\caption{SDSS DR10Q data set fit and extrapolation chi-squares followed by per-observation median chi-squares of extrapolation and associated ratio of outliers.}
\label{tbl:sdss_data set_chi2}
\end{table}	

\begin{figure}
\includegraphics{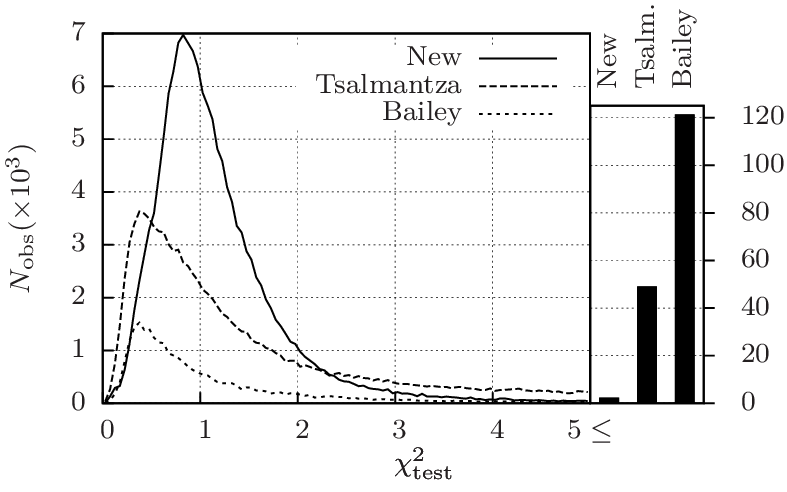}
\caption{Distribution of the observation's $\chisqtest$. Right-hand part represents the number of QSOs ($\times10^{3}$) having $\chisqtest \geq 5$.}
\label{fig:comparison_sdss_detail}
\end{figure}	
	
	We see that $\chisqfit$ is quite stable for all the algorithms while being a bit higher for our algorithm as expected. Now watching at the extreme differences of $\chisqtest$ for the various algorithms and given the fact that we know it to be an estimator that is quite sensitive to outliers, we found relevant to see how these discrepancies are distributed among the observations. To this aim, we computed a $\chisqtest$ per-observation whose distribution is summarized in Fig. \ref{fig:comparison_sdss_detail}.

	Given the significant number of observations used in this test, we can already draw some general trends. Beside the fact that the other algorithms have a better $\chisqfit$, they often fail to satisfactorily extrapolate the spectra. Indeed, for the Bailey algorithm, $81\%$ of observations have $\chisqtest \geq 5$ with a peak up to $9.59 \times 10^{15}$ (median=$8.24 \times 10^{4}$); the Tsalmantza algorithm has respectively $33\%$ of observations having $\chisqtest \geq 5$ and a peak up to $5.78 \times 10^{8}$ (median=$1.789$) while our new algorithm has only $1.4\%$ of `outliers' with a maximal peak of $172$ (median=$1.021$). If the observations are individually compared, our algorithm has a lower $\chisqtest$ in $90\%$ of the time regarding Bailey algorithm and $68\%$ of the time regarding Tsalmantza algorithm while in other cases differences remain quite moderate with a median $\Delta\chisqtest \approx 0.3$. For completeness, we have to note that these huge discrepancies can be mainly attributed to the large amount of missing data ($76\%$) within the resulting rest-frame DR10Q data set.
	
	Another quality one would often desire is the ability to have the most general and representative set of principal components able to model unseen observations. In this optic, two tests were performed. 

\begin{figure*}
\includegraphics{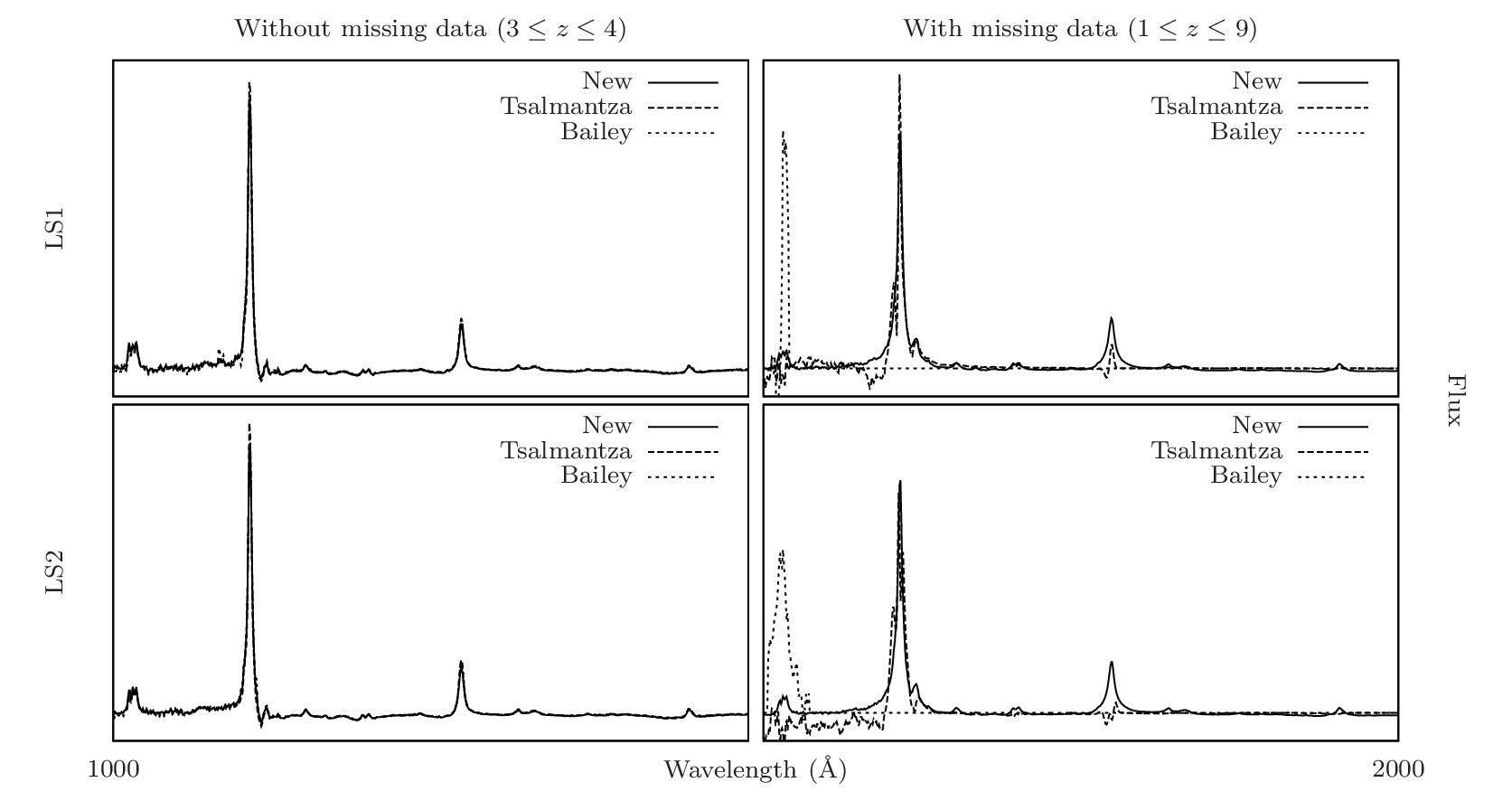}
\caption{Principal components retrieved from two independent data sets (LS1 and LS2) coming from the DR10Q catalogue. The left-hand part corresponds to the dominant eigenvectors extracted from spectra having no missing data while the right-hand part are the associated components coming from spectra having missing fluxes.}
\label{fig:comparison_ls1_ls2}
\end{figure*}	
	
	First, we split the initial DR10Q data set into two subsamples each of 74 025 spectra spanning from 4000 to 10 000\r{A} (hereafter LS1 and LS2). From each of these subsamples, we extract $\sim$8600 spectra for which the rest-frame region from 1000 to 2000\r{A} is entirely covered -- and consequently `without' missing data -- and $\sim$62 000 spectra for which the mentioned region is only partially covered -- and thus `with' missing data. Note that according to the selected observed wavelengths and the previous definitions, spectra without missing data will correspond to those having $3 \leq z \leq 4$ and spectra with missing data to those having $1 \leq z \leq 9$.
	
	The consistency and convergence of each algorithm were then tested by retrieving 10 principal components out of LS1 and LS2 data sets taken with and without missing data. Prior to discussing the results, we have to mention that the components may be swapped from data sets with missing data to the corresponding ones without missing data. This is easily explained by the fact that some patterns may be highlighted due to the uneven coverage of the wavelength range while being damped in the total variance if a full coverage is considered. Fig. \ref{fig:comparison_ls1_ls2} illustrates the results of this test regarding the first component of the data set without missing data and associated component with missing data. We see that in absence of missing data, all algorithms are consistent and succeed in converging towards the dominant eigenvector while in the case of missing data only our algorithm shows both a good consistency and a good convergence. More precisely, regarding our algorithm, the mean differences per-point between LS1 and LS2 are in $\bigo{10^{-6}}$ for both subsets with and without missing data. Differences noticed between subsets, that is mainly the general larger equivalent width and the stronger \ion{N}{V}{1240}\r{A} emission line of the subset with missing data, are consistent with the underlying data set and make it able to model the larger variety of spectra coming from the larger redshift range. Let us mention, that no significant inconsistencies have been noticed up to the sixth component for LS1 and LS2 on both subsets for our algorithm. Concerning the Bailey's algorithm, we see, as already suggested in Section \ref{sec:comparison_simulated}, that it fails to converge if a large amount of data is missing. Tsalmantza's implementation shows a better convergence but fails to correctly reproduce some emission lines (\ion{O}{VI}{1033}, \ion{N}{V}{1240}, \ion{C}{IV}{1549} and \ion{C}{III}{1908}\r{A}) as well as the region $\leq$1200\r{A} in subsets with missing data. Similarly to Bailey's algorithm, analysis show that these are due to a convergence problem occurring on sparse data sets and arising from numerical instabilities.
		
		In the second test performed, 10 principal components were retrieved from a set of 98 700 spectra spanning from 4000 to 10 000\r{A} (hereafter LS3). We tested the fact that the principal components are the best at individually describing the underlying data variance by fitting the remaining 49 350 spectra (hereafter VS) using a subset of five components retrieved from LS3. That is, if the variance explained by each individual component is maximized then the fit of a smaller number of components to a similar data set should be minimal. In this case, if we consider computing $\chi^2$ of VS data set as defined by equation \eqref{eq:data set_chisq} for each of the algorithms, we have respectively $\chi^2=0.372$ for the Bailey algorithm, $\chi^2=0.316$ for the Tsalmantza algorithm and $\chi^2=0.309$ for our algorithm, thus supporting our last quality criterion. Note again that if all components were considered, Bailey and Tsalmantza algorithms would have had a better resulting $\chi^2$.

\section{Discussion}
\label{sec:discussion}

\subsection{Convergence and uniqueness}
\label{sec:discussion_convergence}
At first glance it is surprising that the power iteration algorithm works at all, but in practice, it is easily demonstrated. Consider a diagonalizable square matrix $\mat{A}$ of size ($n \times n$) having eigenvectors $\vec{p_1}, ..., \vec{p_n}$ and associated eigenvalues $\lambda_1, ..., \lambda_n$ where $\abs{\lambda_1} \geq \abs{\lambda_2} \geq \cdots \geq \abs{\lambda_n}$. Since the eigenvectors are orthogonal between each other, we can write the starting vector of the power iteration algorithm as
\begin{equation}
\vec{u^{(0)}} = c_1 \vec{p_1} + ... + c_n \vec{p_n}. 
\end{equation}
Then we will have that the vector at iteration $k$ given by
\begin{eqnarray}
\vec{u^{(k)}} & = & \mat{A}^k \sum_{i = 1}^n c_i \vec{p_i} = \sum_{i = 1}^n c_i \lambda_i^k \vec{p_i} \nonumber \\
& = & \lambda_1^k \left[ c_1 \vec{p_1}  + \sum_{i = 2}^n c_i \left(\dfrac{\lambda_i}{\lambda_1}\right)^k \vec{p_i} \right].
\end{eqnarray}
We see that $\vec{u^{(k)}}$ will converge to $\lambda_1^k c_1 \vec{p_1}$ as $k \rightarrow \infty$ under the following conditions: $c_1 \neq 0$, i.e. the starting vector has a nonzero component in the direction of the dominant eigenvector and $\abs{\lambda_1} > \abs{\lambda_2}$, i.e. the data set has only one dominant eigenvalue. Also note that, in this case, the rate of convergence to the dominant eigenvector will be principally given by $\lambda_1/\lambda_2$ and that in the case of $\lambda_1 = \lambda_2$, the uniqueness of the solution is not guaranteed as it depends on the starting vector $\vec{u^{(0)}}$. Refinement seen in Section \ref{sec:WPCA_new_refinement} obeys to the same conditions. Additionally, the dominant eigenvalue of $\left(\mat{A} - d \mati\right)^{-1}$ will tend to $\infty$ as $d \rightarrow \lambda$, where $\mat{A}$ is the covariance matrix, $d$ the current `eigenvalue' (i.e. the Rayleigh quotient of $\mat{A}$ and $\vec{u}$, the current `eigenvector') and $\lambda$ a real eigenvalue of $\mat{A}$. This may thus lead to numerical instabilities that can strongly deteriorate the vector used in the next step of the Rayleigh quotient iteration.

	That being said, failures against convergence can be easily checked. For example, a satisfactory solution should have $\mat{A} \vec{p} \approx \lambda \vec{p}$; and if not, the power iteration algorithm can be resumed with another starting vector $\vec{u^{(0)}}$. Under the condition that we have $\lambda_1 = \lambda_2$, we will have an infinite number of eigenvectors that are equally good at describing the data set variance, thus choosing one over another is irrelevant. Concretely, during the $\sim$100 000 tests performed in the context of this paper, no such problems arose. Checks performed on eigenvectors from Section \ref{sec:comparison_simulated} and \ref{sec:comparison_observation} show that the latter are orthogonal to a machine precision of $\bigo{10^{-16}}$ and that the weighted covariance matrix (as described by equation \ref{eq:covar_mat_new}) is diagonalized with the same precision. Nevertheless, people wanting extreme reproducibility and/or secure convergence may still use the SVD in order to extract all eigenvectors from equation \eqref{eq:covar_mat_new} at the expense of a lower flexibility (see Sections \ref{sec:discussion_apriori} and \ref{sec:discussion_smoothing}).

\subsection{A priori eigenvectors}
\label{sec:discussion_apriori}
In some situations, one may already have an approximation of the wanted principal components corresponding to a given data set. Typical examples include principal components update according to new observations added to the data set or in a real case we encountered, the use of SDSS DR9Q eigenvectors to the DR10Q data set. These will constitute a priori eigenvectors that it would be regrettable not to use. The design of our algorithm allows us to easily take benefits of these a priori vectors. Instead of using a random vector, $\vec{u}^{(0)}$, in equation \eqref{eq:power_iteration} one can straightforwardly substitute the known vectors to the random starting vector usually used. Doing so will typically decrease the number of iterations needed to converge towards the new vectors. Beware that vectors may be swapped between a priori eigenvectors and effective data set eigenvectors -- as encountered in Section \ref{sec:comparison_observation} -- and consequently iterations needed to converge will be the same as if we had used a random starting vector. In this case, one can perform no power iteration and use only the refinement seen in Section \ref{sec:WPCA_new_refinement} in order to converge to the nearest eigenvector.

\subsection{Smoothing eigenvectors}
\label{sec:discussion_smoothing}
It sometimes happens for some variables within eigenvectors to exhibit some sharp features that we know to be artefacts. These occur mainly in cases where we have noise that is comparable in amplitude to the data variance, data regions covered only by a few numbers of observations or data sets containing corrupted observations. Again the flexibility of our algorithm allows us to efficiently deal with these drawbacks. Suppose that we retrieved such a `noisy' eigenpair $\pair{\vec{p}}{\lambda}$ from a given data set whose covariance matrix is given by $\mat{A}$, then, before removing the variance in the direction of the found component through equation \eqref{eq:rayleigh_subtract}, one can filter $\vec{p}$ thanks to any existing smoothing function. Assuming that the resulting vector, $\vec{p^{\prime}}$, is near $\vec{p}$ regarding the norm of their difference, we can suppose that the variance accounted for by $\vec{p^{\prime}}$ (i.e. its `eigenvalue') is similar to the one of $\vec{p}$ and then subtract it along the direction of $\vec{p^{\prime}}$ by

\begin{equation}
\mat{A^\prime} = \mat{A} - \lambda \vec{p^{\prime}} \oprod \vec{p^{\prime}}.
\end{equation}

Note that the kind of smoothing function to use is highly data-dependent and no rule of thumb exists. Nevertheless, a quite general and commonly used filter producing efficient results in the field of QSOs can be found in \cite{savitzky1964}. Finally, let us note that, when applying a filter, the orthogonality of the components has to be manually checked and that they will obviously not diagonalize the covariance matrix anymore.

\subsection{Scaling performance}
\label{sec:scaling_performance}

	PCA is often used in cases where we have a lot of observations and a reasonable number of variables, typically we have $\nobs \gg \nvar$. We know, for example, the classical algorithm to require $\bigo{\nvar^3}$ basic operations in order to solve for the eigenvectors of the covariance matrix and $\bigo{\nvar^2\nobs}$ operations to build this matrix. In the following, we will compute the algorithmic complexity of the various explored algorithms in a similar way.
	
	The Tsalmantza algorithm requires for each iteration (hereafter $\niter$) the solution of linear systems of equations for each of the E and M-step, respectively, in each observation and in each variable. We will then have that its algorithmic complexity is given by $\bigo{\niter \ncomp^3 \left( \nobs + \nvar \right)}$. The Bailey algorithm will be identical except for the M-step that will be in $\bigo{\nvar\nobs\ncomp}$ and can thus be discarded, giving $\bigo{\niter \ncomp^3 \nobs}$. Finally, our algorithm mainly requires the building of the covariance matrix, $\niter$ matrix multiplications, potentially $\nrefine$ matrix inversions in order to refine the eigenvectors and a final single E-step similar to the one for the previous algorithm (discarded here), thus giving $\bigo{\nvar^2\nobs+\niter\nvar^2\ncomp+\nrefine\nvar^3}$.
	
	We see that in case $\nobs \gg \nvar$, our algorithm is much faster than the other ones as the computing time is mainly spent in the covariance matrix building. As an illustration, if we take data similar to those described in Section \ref{sec:comparison_simulated} with $\nobs =$ 10 000, PCA retrieval takes $\sim 140$s for the Bailey and Tsalmantza algorithms (with $\niter=100$) and $\sim 3$s for the new one (with $\niter=$10 000) on a 2.4Ghz CPU.
	
\section{Conclusions}
\label{sec:conclusion}
We presented a new method for computing principal components based on data sets having noisy and/or missing data. The underlying ideas are intuitive and lead to a fast and flexible algorithm that is a generalization of the classical PCA algorithm. Unlike existing methods, based on lower-rank matrix approximations, the resulting principal components are not those that aim to explain at best the whole variance of a given data set but rather those that are the most suitable in identifying the most significant patterns out of the data set while explaining most of its variance. The main benefits of the current implementation are a better behaviour in presence of missing data as well as faster run times on data set having a large amount of observations. Privileged problems encompass data set extrapolation, patterns analysis and principal component usage over similar data sets. We assessed the algorithm performance on simulated data as well as on QSO spectra to which many applications are already foreseen in the field of the \textit{Gaia} mission.

\section{Acknowledgements}
The author acknowledges support from the ESA PRODEX Programme `Gaia-DPAC QSOs' and from the Belgian Federal Science Policy Office.

Funding for SDSS-III has been provided by the Alfred P. Sloan Foundation, the Participating Institutions, the National Science Foundation and the U.S. Department of Energy Office of Science. The SDSS-III web site is http://www.sdss3.org/.

SDSS-III is managed by the Astrophysical Research Consortium for the Participating Institutions of the SDSS-III Collaboration including the University of Arizona, the Brazilian Participation Group, Brookhaven National Laboratory, Carnegie Mellon University, University of Florida, the French Participation Group, the German Participation Group, Harvard University, the Instituto de Astrofisica de Canarias, the Michigan State/Notre Dame/JINA Participation Group, Johns Hopkins University, Lawrence Berkeley National Laboratory, Max Planck Institute for Astrophysics, Max Planck Institute for Extraterrestrial Physics, New Mexico State University, New York University, Ohio State University, Pennsylvania State University, University of Portsmouth, Princeton University, the Spanish Participation Group, University of Tokyo, University of Utah, Vanderbilt University, University of Virginia, University of Washington and Yale University.

\newpage
\appendix
\section{Matlab/Octave source code}
\begin{verbatim}
%   Function designed to find the weighted PCA of a
% given data set such that the resulting principal
% components diagonalize the associated weighted 
% covariance matrix and that the associated principal
% coefficients are the linear coefficients of these
% principal components that best match the data set
% in a least-squares sense.
%
% INPUT:
%   X       : data set matrix from which the weighted
%             mean was already subtracted. Rows
%             correspond to variables, columns to
%             observations.
%   W       : Weight matrix (same size as X).
%   ncomp   : Number of principal components to
%             retrieve.
%   niter   : Number of iterations (allows 
%             convergence to the dominant principal
%             component)
%   nrefine : Number of refinments (allows refinement
%             of the dominant principal component)
%   xi      : Regularization factor controlling the
%             influence of rare variables
% OUTPUT:
%   P       : Principal components arranged in columns
%   C       : Principal coefficients such that X = P*C
%             in a least-squares sense
% AUTHOR:
%   Ludovic Delchambre
%   Extragalactic Astrophysics and Space Observations,
%   Institute of Astrophysics and Geophysics,
%   University of Liege, Belgium
function [P, C] = WPCA(X, W, ncomp, niter, nrefine, xi)
  nvar = size(X,1); nobs = size(X,2);
  P=zeros(nvar,ncomp);C=zeros(ncomp,nobs);
  ws = sum(W,2);
  covar = ((ws * ws').^xi).*((X.*W)*(X.*W)')./(W*W');
  covar(isnan(covar)) = 0;
  for i=1:ncomp
    u = ones(nvar,1) / nvar;
    for j=1:niter
      u = covar*u;
      d = sqrt(u'*u);
      u = u / d;
    end
    d = u'*covar*u;
    for j=1:nrefine
      u = inv(covar - d * eye(nvar)) * u;
      u = u / sqrt(u'*u);
      d = u'*covar*u;
    end
    covar = covar - u*d*u';
    P(:,i) = u;
  end
  for i=1:nobs
    w = diag(W(:,i)).^2;
    C(:,i) = inv(P'*w*P) * (P'*w*X(:,i));
  end
end
\end{verbatim}

\label{lastpage}

\end{document}